\documentclass{ws-rv9x6}
\usepackage{graphicx}
\usepackage{ws-rv-van}
\begin{document}
\chapter[First Passage in Conical Geometry and Ordering of Brownian Particles]
{First Passage in Conical Geometry and \\ Ordering of Brownian
Particles}
\author[E.~Ben-Naim and P.~L.~Krapivsky]{E. Ben-Naim${}^1$ and P.~L.~Krapivsky${}^2$}
\address{${}^1$Theoretical Division and Center for Nonlinear
Studies, Los Alamos National Laboratory, Los Alamos, New Mexico
87545, USA}
\address{${}^2$Department of Physics,
Boston University, Boston, Massachusetts 02215, USA}
\begin{abstract}
We survey recent results on first-passage processes in unbounded cones
and their applications to ordering of particles undergoing Brownian
motion in one dimension. We first discuss the survival probability
$S(t)$ that a diffusing particle, in arbitrary spatial dimension,
remains inside a conical domain up to time $t$. In general, this
quantity decays algebraically $S\sim t^{-\beta}$ in the long-time
limit. The exponent $\beta$ depends on the opening angle of the cone
and the spatial dimension, and it is root of a transcendental equation
involving the associated Legendre functions.  The exponent becomes a
function of a single scaling variable in the limit of large spatial
dimension.  We then describe two first-passage problems involving the
order of $N$ independent Brownian particles in one dimension where
survival probabilities decay algebraically as well. To analyze these
problems, we identify the trajectories of the $N$ particles with the
trajectory of one particle in $N$ dimensions, confined to within a
certain boundary, and we use a circular cone with matching solid angle
as a replacement for the confining boundary.  For $N=3$, the confining
boundary is a wedge and the approach is exact. In general, this ``cone
approximation'' gives strict lower bounds as well as useful estimates
for the first-passage exponents.  Interestingly, the cone
approximation becomes asymptotically exact when $N\to\infty$ as it
predicts the exact scaling function that governs the spectrum of
first-passage exponents.
\end{abstract}

\section{Introduction}

Triggering of earthquakes as stress surpasses a threshold \cite{mo},
onset of avalanches as mass accumulates beyond stability \cite{pb},
outbreak of an epidemic once the number infected individuals exceeds a
threshold \cite{jdm}, and execution of financial transactions once
stock prices reach a prescribed value \cite{bp}: All are vivid
examples of first-passage processes. Indeed, first-passage processes
are ubiquitous and are becoming increasingly relevant for
understanding stochastic processes in the physical and the natural
sciences \cite{wf,ngv,sr}.

A first-passage process characterizes how a stochastic variable
reaches a fixed threshold for the {\em first time}. The survival
probability $S(t)$ that the threshold is not reached up to time $t$ is
a central quantity in the theory of first-passage processes
\cite{wf,sr}. Basic characteristics follow directly from the survival
probability, e.g., $\langle t\rangle=-\int_0^\infty dt\, t\, (dS/dt)$
measures the average duration of the process; this quantity may be
finite or infinite.  The first-passage probability that the threshold
is ever reached equals $1-S(\infty)$; when $S(\infty)>0$, the
threshold may never be reached.

The widely-known gambler ruin problem which involves the location of a
simple random walk on a line is a classic first-passage problem. This
process ends when the wealth of the gambler reaches either zero or a
high threshold \cite{wf,sr}. First-passage properties of multiple
random walks \cite{ghw,hcb} and particles undergoing Brownian motion
\cite{bdup,mp,book} in one dimension are the subject of ongoing
research and underlie dynamics of spin chains \cite{dhp,msbc,bms},
voting processes \cite{kbr,hg,hn}, urn model \cite{abk}, and
unraveling of knots\cite{bdve}.  Many of these first-passage processes 
are closely related to the ordering of a set of independent Brownian
particles on a line \cite{mef,hf,fg,bg,djg,kr,dba,bjmkr,ck,dlb,yal,bmr}.

When the number of random walks is no larger than three, first-passage
processes involving the order of the particles map onto diffusion in a
two-dimensional wedge \cite{mef,dba,bjmkr,sr}. For an arbitrary number
of particles, a similar mapping remains useful. We recently showed
that first-passage processes in high-dimensional cones are connected
with a host of problems involving the order of $N$ Brownian particles
on a line \cite{bk1,bk2,eb}. In this article, we review these results.

We first discuss the survival probability $S(t)$ that a particle
diffusing in an infinite circular $d-$dimensional cone with opening
angle $\alpha$ does not cross the cone boundary during the time
interval $(0,t)$\cite{rdd,dz,bs,bds,ntv,jcm,mkk,leld}.  The survival
probability decays algebraically with time
\begin{equation}
\label{S-decay}
S\sim t^{-\beta}\quad \text{as}\quad t\to\infty. 
\end{equation}
Since $S(\infty)=0$, the surface of the cone is eventually
reached. The exponent $\beta$ turns out to be a root of a
transcendental equation involving the associated Legendre functions.
We also compute the mean first-passage time which is finite when
$\beta>1$.

In principle, the first-passage exponent $\beta$ depends on the
opening angle $\alpha$ and the spatial dimension $d$. For $d\gg 1$,
the exponent becomes a function of a single scaling variable \cite{gib}
which is a special combination of $\alpha$ and $d$.  The exponent
$\beta$ is root of the equation \cite{bk1}
\begin{equation}
\label{scaling1}
D_{2\beta}(y)=0\quad {\rm with}\quad y=(\cos\alpha)\sqrt{d},
\end{equation}
involving the Parabolic cylinder function $D_\nu$ \cite{NIST}. This
scaling behavior shows that $\beta$ is of order unity only when the
cone is nearly flat (close to a half space).  We also discuss a
variety of asymptotic properties, e.g., the behavior in the interior
and the exterior of very thin cones.

Next, we discuss first-passage processes involving the order $N$
independent Brownian particles in one dimension. We tag the particles
according to initial position.  Regardless of the initial conditions,
the order of the particles eventually gets ``scrambled'' and since
diffusion is an ergodic process, all $N$ permutations of particle
order become equally probable.  We show that there is a host of
interesting first-passage questions that characterize how the initial
order unravels.

For example, we rank the particles from right to left according to
initial position with rank $n=1$ corresponding to the rightmost
particle (the ``leader'') and rank $n=N$ to the leftmost particle (the
``laggard''). We ask: what is the survival probability $R_n(t)$ that
the rank of the initial leader at time $t=0$ does not fall below $n$
until time $t$.  The quantity $R_1$ gives the probability that the
leader kept the lead, while $R_{N-1}$ is the probability that the
leader is never last \cite{bg,kr,bjmkr}.  In general, the survival
probabilities decay algebraically,
\begin{equation}
\label{R-decay}
R_n\sim t^{-\gamma_n},
\end{equation}
in the long-time limit. Interestingly, there is a family of $N-1$
distinct first-passage exponents with
$\gamma_1>\gamma_2>\cdots>\gamma_{N-1}$.

\begin{figure}[t]
\centerline{\includegraphics[width=0.5\textwidth]{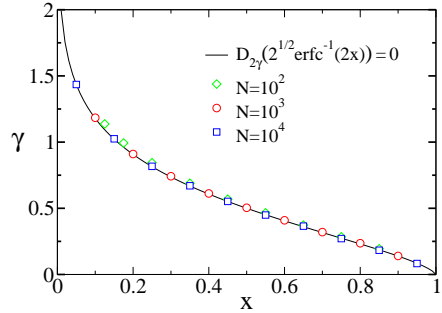}}
\caption{The exponent $\gamma$ versus the scaling variable $x$. 
Results of Monte Carlo simulations with $N=10^2$, $N=10^3$, and $N=10^4$ 
particles, are compared with the theoretical prediction \eqref{scaling2}.}  
\label{fig-gamma}
\end{figure}

The trajectories of the $N$ one-dimensional Brownian particles can be
identified with that of a single compound particle undergoing ordinary
Brownian motion in $N$ dimensions. Then $R_n$ is the probability that
this particle remains within some domain in this higher-dimensional
space. When $N\leq 3$, this domain is always a wedge, and the
first-passage exponents can be obtained analytically.  For arbitrary
$N$, we approximate this domain using an unbounded cone that encloses
an identical solid angle. This approach gives a useful approximation
for the decay exponents $\gamma_n$.  When the number of particles is
very large, we exploit the scaling relationship \eqref{scaling1} and
find that the exponents $\gamma_n$, which depend on the threshold $n$
and the number of particles $N$, become a function of a single scaling
variable $n/N$. The exponent $\gamma$ is root of the transcendental
equation \cite{bk2}
\begin{equation}
\label{scaling2}
D_{2\gamma}\left(\sqrt{2}\,{\rm erfc}^{-1}(2x)\right)=0\qquad {\rm with}
\qquad x=n/N,
\end{equation}
involving the parabolic cylinder function and the inverse
complementary error function. As shown in figure \ref{fig-gamma},
numerical simulations reveal that this result is asymptotically exact:
The scaling exponent $\gamma_n$ is a function of the scaling variable
$x=n/N$, and the scaling function is given by \eqref{scaling2}.

The rest of this paper is organized as follows. In section 2, we
provide a pedagogical discussion of first-passage in a two-dimensional
wedge. We derive the first-passage exponent directly from the
diffusion equation governing the survival probability and show that
problems involving particle order map onto a two dimensional wedge
when $N=3$. We then generalize this theoretical description to
arbitrary spatial dimensions and obtain the first-passage exponent in
section 3. Relevant asymptotic relations including in particular the
scaling function underlying $\beta$ in the infinite-dimension limit
are obtained in section 4. In section 5, we develop the cone
approximation, using a three-dimensional corner geometry as an
illustrative example.  We then discuss rank statistics and use the
cone approximation to estimate the first-passage exponents and to
derive their asymptotic behavior (section 6). In section 7, we
introduce the number of pair inversions as a measure of order. Again,
we find that the cone approximation yields useful estimates for the
first-passage exponents and that it becomes asymptotically exact when
the number of particles diverges. We conclude in section 8.

\section{Wedges} 

\begin{figure}[t]
\centerline{\includegraphics[width=0.3\textwidth]{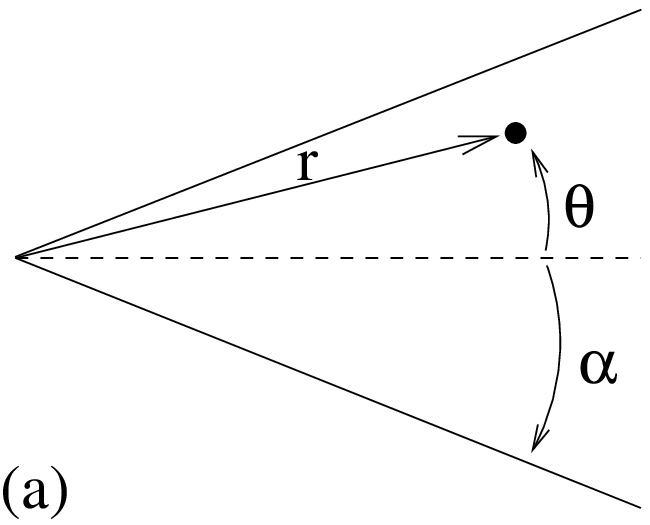}\hspace{.5in}
\includegraphics[width=0.3\textwidth]{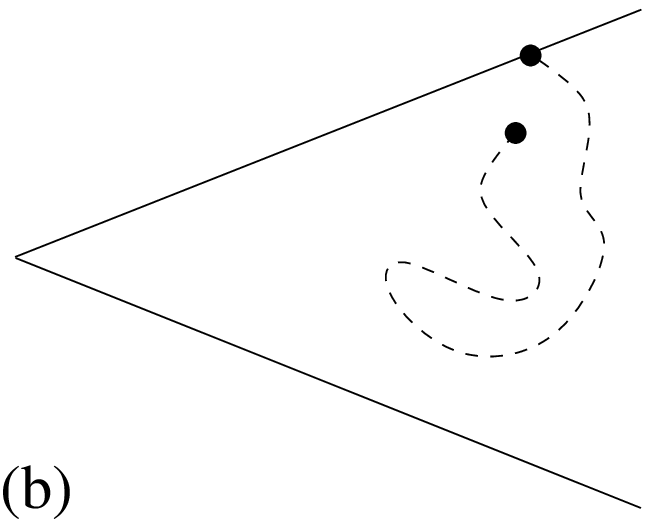}}
\caption{(a) A wedge with opening angle $\alpha$. The radial
coordinate $r$ and the polar coordinate $\theta$ specify the initial
location of the Brownian particle. (b) Illustration of the
first-passage process: the trajectory of a Brownian particle is
indicated by a dashed line while the start and end points are
indicated by bullets.}
\label{fig-wedge}
\end{figure}

Consider a particle undergoing simple diffusion in a wedge
(Fig.~\ref{fig-wedge}a) with an opening angle $\alpha$. The angle
$\alpha$ is measured from the center of the wedge and hence
$\alpha\leq \pi$.  The first-passage process ends when the particle
reaches the boundary of the wedge (Fig.~\ref{fig-wedge}b).

The survival probability that the particle avoids the wedge boundary
up to time $t$ depends on its initial location $(r,\theta)$ of the
particle: $S\equiv S(r,\theta,t)$ where $r$ is the distance to the
cone apex and $\theta$ is the polar angle
(Fig.~\ref{fig-wedge}a). Since the particle undergoes simple Brownian
motion, the survival probability obeys the diffusion equation
\cite{ghw}
\begin{equation}
\label{S-eq}
\frac{\partial S(r,\theta,t)}{\partial t}=D\nabla^2 S(r,\theta,t),
\end{equation}
where $D$ is the diffusion coefficient. The boundary condition
$S(r,\alpha,t)=0$ reflects that the first-passage process ends when
the particle reaches the boundary, and the initial condition is
$S(r,\theta,t=0)=1$ for all $\theta<\alpha$.

The survival probability decays algebraically 
\begin{equation}
\label{decay}
S(r,\theta,t) \simeq \,\Phi(r,\theta)\, t^{-\beta},
\end{equation}
in the long-time limit. The algebraic dependence on time is well-known
as for $\alpha=\frac{\pi}{2}$ the problem is effectively
one-dimensional and $\beta=\frac{1}{2}$. Here, we show how the decay
exponent $\beta$ can be conveniently derived from \eqref{S-eq}. By
substituting the algebraic decay \eqref{decay} into the diffusion
equation \eqref{S-eq}, we conclude that the amplitude $\Phi$ obeys
Laplace's equation
\begin{equation}
\label{Phi-eq}
\nabla^2 \Phi(r,\theta)=0.
\end{equation}
The boundary conditions are 
\hbox{$\Phi(r,\alpha)=\Phi(0,\theta)=0$}. The amplitude must be
positive inside the wedge, $\Phi(r,\theta)>0$ for $\theta<\alpha$. The
Laplace equation \eqref{Phi-eq} manifests the direct connection
between diffusive first-passage processes and electrostatics
\cite{hcb,jdj}.

Since the quantity $S$ is dimensionless, and since the only
dimensionless combination of the quantities $r$, $D$, and $t$ is
$Dt/r^2$, the amplitude necessarily has the form
\begin{equation}
\label{Phi-sep}
\Phi(r,\theta)=\left(\frac{r^2}{D}\right)^{\beta}\psi(\theta).
\end{equation}
The function $\psi\equiv \psi(\theta)$ depends on the polar angle
$\theta$ alone. This function is positive inside the wedge, and it
vanishes on the boundary, $\psi(\alpha)=0$. We now substitute
\eqref{Phi-sep} into \eqref{Phi-eq} and use
\begin{equation*}
\label{lap-2}
\nabla^2 \equiv \frac{\partial^2}{\partial r^2}+
\frac{1}{r}\frac{\partial}{\partial r}+
\frac{1}{r^2}\frac{\partial^2}{\partial \theta^2}
\end{equation*}
to find that $\psi(\theta)$ obeys the simple eigenvalue equation
\begin{equation}
\label{psi-eq-2}
\frac{\partial^2\psi}{\partial \theta^2}+(2\beta)^2\psi=0.
\end{equation}
Equation \eqref{psi-eq-2} involves only the angular component of the
Laplacian.  By symmetry, we expect that the function $\psi(\theta)$ is
even, $\psi(\theta)=\psi(-\theta)$. The solution to 
Eq.~\eqref{psi-eq-2} is $\psi(\theta)=\cos(2\beta\theta)$, and the 
first-passage exponent \cite{S58}
\begin{equation}
\label{beta-2}
\beta=\frac{\pi}{4\alpha}
\end{equation}
follows immediately from boundary condition $\psi(\alpha)=0$. We see
that the asymptotic behavior \eqref{decay} with the exponent
\eqref{beta-2} holds for all initial conditions. The initial location
affects only the amplitude $\Phi(r,\theta)$. Furthermore, using
\eqref{decay} and \eqref{Phi-sep} we conclude that $S \sim
r^{2\beta}$.

The exponent is minimal for a needle, $\beta(\pi)=1/4$ \cite{cr}, and
hence, in two dimensions the particle is bound to encounter a
semi-infinite needle \cite{cr}. The exponent $\beta(\alpha)$ diverges
for infinitely thin wedges, and the duration of the first-passage
process is infinitesimal in the limit $\alpha\to 0$. Finally, since
$\beta=1$ when $\alpha=\pi/4$, the mean first-passage time is finite
if and only if the wedge is smaller than a square corner.

\begin{figure}[t]
\centerline{\includegraphics[width=0.35\textwidth]{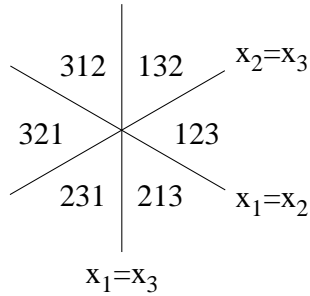}}
\caption{A projection of three dimensional space onto the plane
perpendicular to the diagonal line $x_1=x_2=x_3$ divides space into
six equal wedges, each with distinct particle order.}
\label{fig-pie}
\end{figure}

First-passage problems involving the order of three particles in
one-dimension map onto a wedge.  Let $x_i(t)$ be the location of the
$i$th particle at time $t$. Without loss of generality, we label the
particles according to the initial positions,
$x_1(0)>x_2(0)>x_3(0)$. Furthermore, we can view the three coordinates
$(x_1,x_2,x_3)$ as those of a ``composite'' particle in three
dimensions. Since each particle undergoes independent Brownian motion,
the composite particle undergoes Brownian motion in three
dimensions. The three planes $x_1=x_2$, $x_1=x_3$, and $x_2=x_3$,
divide space into six equal wedges \cite{bjmkr,de} with opening angle
$\alpha=\pi/6$ (figure \ref{fig-pie}), and in each wedge the particle
order is distinct.

\begin{figure}[t]
\centerline{\includegraphics[width=0.3\textwidth]{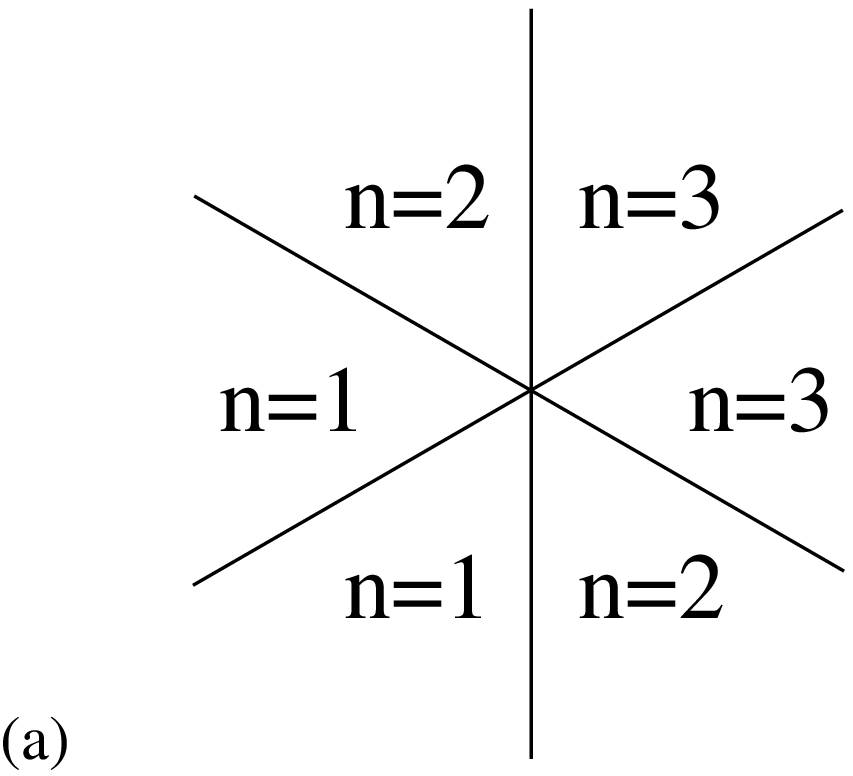}\hspace{.75in}\includegraphics[width=0.3\textwidth]{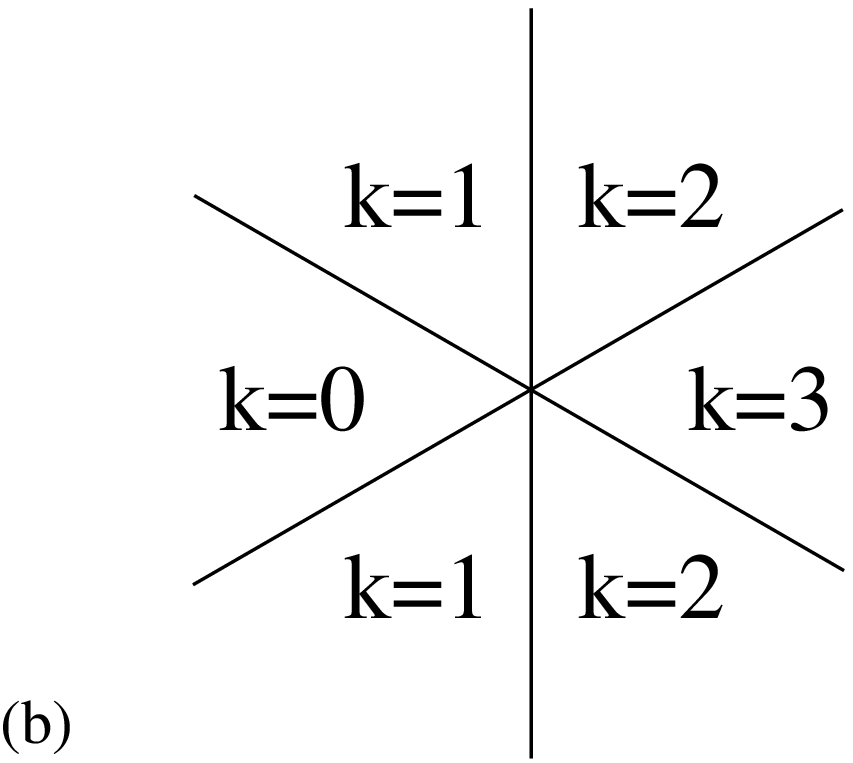}}
\caption{(a) The rank $n$ of the initial rightmost particle in each of
the six wedges shown in figure \ref{fig-pie}. (b) The number of
inversion $k$ in each of the six wedges.}
\label{fig-six}
\end{figure}

Consider the rank of the initial rightmost particle. Let
$R_n$ be the survival probability that this rank does not fall below
$n$ up to time $t$. Figure \ref{fig-six}a shows the rank of the leader
in each of the six wedges.  In two of the wedges $n\leq 1$ and in four
of them $n\leq 2$. Hence, the survival probabilities $R_1$ and $R_2$
are equivalent to the probability that a diffusing particle remains
inside a wedge with opening angles $\alpha=\pi/3$ and $\alpha=2\pi/3$
respectively. Immediately, we obtain the asymptotic behaviors
\begin{equation}
R_1\sim t^{-3/4}\qquad{\rm and}\qquad R_2\sim t^{-3/8}.
\end{equation}
Using the definition \eqref{R-decay}, the set of exponents is
$\{\gamma_1,\gamma_2\}=\{3/4,3/8\}$. Of course, if there are only two
particles, the problem is trivially equivalent to a wedge with opening
angle $\alpha=\pi/2$ and the first passage exponent is
$\gamma_1=1/2$. Based on these two examples, we expect that generally
there are $N-1$ exponents $\gamma_n(N)$ that obey
$\gamma_1>\gamma_2>\cdots>\gamma_{N-1}$.

Figure \ref{fig-pie} shows that there is a set of six first-passage
exponents
\begin{equation}
\label{betal}
\beta_l=\frac{3}{2\,l},
\end{equation}
where $l=1,2,\ldots,6$ is the number of wedges.  Specifically, the set
is $\{1/4,3/10,3/8,1/2,3/4,3/2\}$, although for the rank problem,
only two of these values are realized.

There are additional measures of particle order, each with a different
first-passage process. The rank characterizes only the location of one
particle with respect to the rest of the particles. The number of pair
inversions, $k$, measures the number of pairs of particles that are
inverted with respect to the initial configuration \cite{mb,dek}.  For
example, if the order changes from $321$ initially to $312$ later,
then number of inversions increases from $k=0$ to $k=1$ because the
pair $12$ is inverted with respect to the initial configuration. The
number of inversions is maximal, $k=3$, when the particle order is
completely reversed ($321\to 123$) and hence $0\leq k\leq 3$.

What is the survival probability $M_k$ that the number of
pair inversions remains smaller than $k$ up to time $t$? When $k=0$,
the trajectory of the composite particle is confined to a single
wedge, $l=1$. Similarly, the condition $k\leq 1$ corresponds to $l=3$,
and the condition $k\leq 2$ corresponds to $l=5$. Using equation
\eqref{betal}, we find
\begin{equation}
\label{nu3}
M_1\sim t^{-3/2},\qquad M_2\sim t^{-1/2}, \qquad M_3\sim t^{-3/10}.
\end{equation}
In section 7, we show that the number of distinct exponents grows {\em
quadratically} with the number of particles.

\section{Circular Cones} 

\begin{figure}[t]
\centerline{\includegraphics[width=0.35\textwidth]{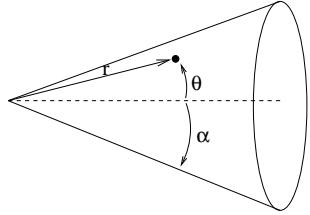}}
\caption{A cone with opening angle $\alpha$.} 
\label{fig-cone}
\end{figure}

An infinite circular cone in arbitrary spatial dimension $d>1$ is 
specified by the opening angle $\alpha<\pi$, that is, the angle
between the cone surface and its axis (see figure \ref{fig-cone}). The
exterior of a cone with opening angle $\alpha$ is a cone with opening
angle $\pi-\alpha$. In all dimensions, a cone with opening angle
$\alpha=\pi/2$ is a half space.

The survival probability of a diffusing particle again depends only
on two variables: the radial coordinate $r$ and the polar angle $\theta$
(see figure \ref{fig-cone}). The survival probability obeys the
diffusion equation \eqref{S-eq} with the Laplacian
\begin{equation}
\nabla^2 \equiv \frac{\partial^2}{\partial r^2}+
\!\frac{d\!-\!1}{r}\frac{\partial}{\partial r}+
\frac{1}{r^2(\sin\theta)^{d-2}}\frac{\partial}{\partial
\theta}(\sin\theta)^{d-2} \frac{\partial}{\partial\theta}.
\end{equation}
In general, the survival probability decays algebraically as in
\eqref{decay} and the amplitude obeys the Laplace equation
\eqref{Phi-eq}. By following the steps leading to \eqref{psi-eq-2} we
find that the angular function $\psi$ satisfies the eigenvalue equation 
\begin{equation}
\label{psi-theta-eq}
\frac{1}{(\sin\theta)^{d-2}}\frac{d}{d \theta}\left[(\sin\theta)^{d-2}
\frac{d\psi}{d  \theta}\right]+2\beta(2\beta+d-2)\psi=0. 
\end{equation}
To solve this equation, we introduce the variable
$\mu=\cos\theta$, and with this transformation, the function
$\psi\equiv \psi(\mu)$ satisfies 
\begin{equation}
\label{psi-mu-eq}
(1-\mu^2)\frac{d^2\psi}{d\mu^2}-(d-1)\,\mu\,\frac{d\psi}{d\mu}
+2\beta(2\beta+d-2)\psi=0.
\end{equation}
Again, the boundary condition is $\psi(\cos\alpha)=0$.

For $d=3$, the solution to Eq.~\eqref{psi-mu-eq} is given by the
Legendre function, \hbox{$\psi_3(\theta)= P_{2\beta}(\cos\theta)$}
\cite{NIST}. The boundary condition implies that the first-passage
exponent is the smallest root of the transcendental equation
\begin{equation}
\label{beta3}
P_{2\beta}(\cos\alpha)=0.
\end{equation} 
Henceforth, we always choose the smallest root (or equivalently,
eigenvalue) because the function $\psi(\theta)$ must remain
positive. Further, equation \eqref{psi-mu-eq} shows that there is a
direct connection between the first-passage exponent $\beta$ and the
lowest eigenvalue of the {\em angular} part of the Laplace operator.

\begin{figure}[t]
\centerline{\includegraphics[width=0.5\textwidth]{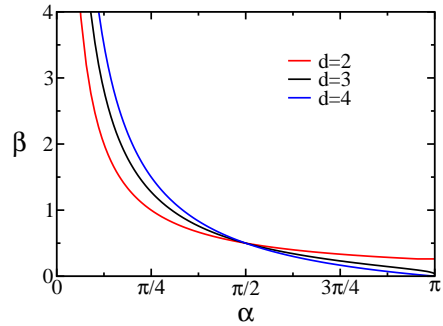}}
\caption{The first-passage exponent $\beta$ versus the opening angle 
$\alpha$ in dimensions $d=2$, $3$, and $4$.}
\label{fig-beta}
\end{figure}

When $d=4$, we can obtain the first-passage exponent explicitly. Using
the transformation $\psi_4(\theta)=(\sin\theta)^{-1}u(\theta)$, the
eigenvalue equation \eqref{psi-theta-eq} becomes
$u_{\theta\theta}+(2\beta+1)^2u=0$. There are two independent
solutions, $\sin[(2\beta+1)\theta]$ and $\cos[(2\beta+1)\theta]$ but
since the function $\psi$ must remain finite, only the former is
physical. Therefore
$\psi_4=\sin\left[(2\beta+1)\theta\right]/\sin\theta$ and the boundary
condition $\psi(\alpha)=0$ gives the exponent
\begin{equation}
\label{beta4}
\beta_4(\alpha)=\frac{\pi-\alpha}{2\alpha}\,.
\end{equation}
In contrast with equation \eqref{beta-2}, the exponent now vanishes
when  $\alpha = \pi$. Hence, in
dimensions three and higher, a needle is not reached 
(Fig.~\ref{fig-beta}). In the limit of very thin cones, the exponent
diverges and is inversely proportional to $\alpha$ as in
\eqref{beta-2}. Equations \eqref{psi-mu-eq}--\eqref{beta3} 
were recently used for analyzing entropic forces
for polymers in the vicinity of a conical tip \cite{mkk}.

To solve the eigenvalue equation \eqref{psi-mu-eq}, we make the transformation 
\begin{equation}
\psi(\mu)=\left(1-\mu^2\right)^{-\delta/2}
\Psi(\mu) \qquad {\rm with}\qquad\delta=\frac{d-3}{2}.
\end{equation}
The auxiliary function $\Psi(\mu)$ obeys
\begin{eqnarray}
\label{leg-eq}
(1-\mu^2)\frac{d^2\Psi}{d\mu^2}&-&2\,\mu\,\frac{d\Psi}{d\mu}
+\left[(2\beta+\delta)(2\beta+\delta+1)-\frac{\delta^2}{1-\mu^2}\right]
\Psi=0.
\end{eqnarray}
This equation specifies the associated Legendre functions
$P_{2\beta+\delta}^\delta(\mu)$ and $Q_{2\beta+\delta}^\delta(\mu)$ of
degree $2\beta+\delta$ and order $\delta$ \cite{NIST}. Since $\psi$
must remain finite when $\alpha\to 0$, the first (second) solution is physical
when $d$ is odd (even):
\begin{equation}
\label{psid}
\psi_d(\theta)=
\begin{cases}
(\sin\theta)^{-\delta}P_{2\beta+\delta}^\delta(\cos\theta)& \qquad d~~ {\rm odd},\\
(\sin\theta)^{-\delta}Q_{2\beta+\delta}^\delta(\cos\theta)& \qquad d~~ {\rm even}.
\end{cases}
\end{equation}
Using the fact that the survival probability vanishes on the surface
of the cone, we find that $\beta$ is related to the smallest root of
the associated Legendre function
\begin{equation}
\label{betad}
\begin{split}
P_{2\beta+\delta}^\delta(\cos\alpha) = 0&\qquad d~~ {\rm odd},\\
Q_{2\beta+\delta}^\delta(\cos\alpha) = 0&\qquad d~~ {\rm even}.
\end{split}
\end{equation}
In general, the first-passage exponent $\beta\equiv \beta_d(\alpha)$
depends on the opening angle $\alpha$ and the dimension $d$. The
exponent $\alpha$ decreases monotonically as the opening angle
$\alpha$ increases (figure \ref{fig-beta}).

The solution $\psi_d(\mu)$ is polynomial in $\mu$ when the
first-passage exponent is half integer. For example, $\psi_d(\mu)=\mu$
when $\beta=1/2$ and and $\psi_d(\mu)=d\mu^2-1$ when $\beta=1$.
Hence, the mean-first passage time is finite if only if the opening
angle is small enough $\cos\theta>1/\sqrt{d}$.

In general, the mean first-passage time $T(r,\theta)$ for a particle
with initial coordinates $(r,\theta)$ obeys the Poisson equation
\cite{sr,mmmo}
\begin{equation}
\label{T-eq}
D\nabla^2 T(r,\theta) = -1,
\end{equation}
and satisfies the boundary conditions
$T(r,\alpha)=T(0,\theta)=0$. Dimensional analysis implies that
$T(r,\theta)=(r^2/D)U(\theta)$ with $U$ a dimensionless function of
the angle $\theta$. The first-passage time can be obtained explicitly
\begin{equation}
\label{T}
T(r,\theta)=\frac{r^2}{2D}\,
\frac{\cos^2\theta-\cos^2\alpha}{d\cos^2\alpha-1}.
\end{equation}
This equation holds for $\cos\theta>1/\sqrt{d}$ where $\beta>1$. As
expected, the duration of the first-passage process is a monotonic
function of both distance $r$ and angle $\theta$.

\section{Asymptotic properties}

Our goal is to apply the first-passage results in unbounded cones to
ordering of Brownian particles. We are especially interested in the
behavior of a large number of particles, which requires asymptotic
properties of \eqref{betad} when $d\to\infty$. These asymptotics are
summarized in this section.

Taking the limit $d\to\infty$ while keeping the angle fixed shows that 
\begin{equation}
\label{three}
\lim_{d\to\infty}\beta_d(\alpha)=
\begin{cases}
\infty&\qquad\alpha<\pi/2,\\
1/2&\qquad\alpha=\pi/2,\\
0&\qquad\alpha>\pi/2.\\
\end{cases}
\end{equation}
Hence, the first-passage process is infinitesimally short when the
opening angle is acute, and conversely, it lasts forever when the
angle is obtuse. For a right-angle, $\alpha=\pi/2$, the problem is
effectively one-dimensional. Further, one can also show \cite{bk1}
that the exponent grows linearly $\beta\sim d$ for acute angles,
$\alpha<\pi/2$, while in the complementary case, the exponent decays
exponentially, $\beta\sim (\sin\alpha)^d$.

The limiting behavior \eqref{three} suggests that we
should focus on a shrinking region near $\alpha=\pi/2$. Moreover, the
fact that $\beta=1$ when $\sqrt{d}\cos\alpha=1$ indicates that the
first-passage exponent has the scaling form
\begin{equation}
\label{beta-y-scaling}
\beta_d(\alpha)\to \beta(y) \qquad {\rm with}\qquad y=(\cos\alpha)\sqrt{d},
\end{equation}
when $d\to\infty$. If we take the limits $\mu\to 0$ and $d\to\infty$
in such a way that the scaling variable $z=\mu\sqrt{d}$ is finite,
then with this scaling transformation equation \eqref{psi-mu-eq}
becomes
\begin{equation}
\label{psi-z-eq}
\psi_{zz}-z\,\psi_z+2\beta \psi=0.
\end{equation}
The boundary condition is $\psi(y)=0$. We now make the transformation 
$\psi(z)=\exp(z^2/4)u(z)$ and arrive at the parabolic cylinder equation 
\begin{equation}
u_{zz}+\left(2\beta+\frac{1}{2}-\frac{z^2}{4}\right)u=0.
\end{equation}
The two independent solutions are the parabolic cylinder functions
$D_\nu(z)$ and $D_\nu(-z)$ \cite{NIST}. Using the asymptotic behavior
\hbox{$D_\nu(-z)\sim \exp(z^2/4)$} as \hbox{$z\to\infty$} and the fact
that the survival probability must be finite on the cone axis, we
eliminate the latter solution and hence,
\begin{equation}
\label{psi-z}
\psi(z)=e^{z^2/4}D_{2\beta}(z).
\end{equation}
By invoking the boundary condition $\psi(\mu)=0$ we arrive at our main
result \eqref{scaling1}. We choose the largest root of the parabolic
cylinder function because $\psi$ is positive. The scaling form
\eqref{beta-y-scaling} implies that $\beta$ is finite in the narrow
region of opening angles
\begin{equation}
\alpha-\pi/2\sim d^{-1/2}.
\end{equation}
Hence, the first-passage exponent is finite only when the cone is
sufficiently close to a half space.

\begin{figure}[t]
\centerline{\includegraphics[width=0.5\textwidth]{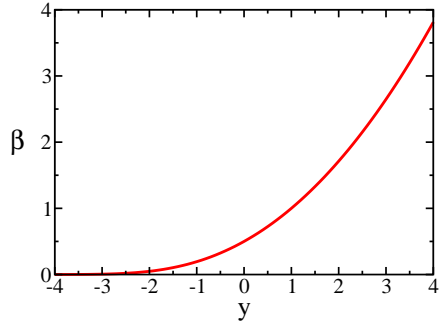}}
\caption{The exponent $\beta$ versus the scaling variable
$y=(\cos\alpha)\sqrt{d}$.}
\label{fig-scaling}
\end{figure}

Asymptotic analysis of Eq.~\eqref{psi-z} reveals the limiting
behaviors (Fig.~\ref{fig-scaling})
\begin{equation}
\label{beta-y-limits}
\beta(y)\simeq
\begin{cases}
\sqrt{y^2/8\pi}\,\exp\!\left(-y^2/2\right)&\qquad y\to-\infty,\\
y^2/8 & \qquad y\to\infty.
\end{cases}
\end{equation}

We also mention the asymptotic behaviors in the interior and the
exterior of a slender cone (the dimension $d$ is fixed)
\begin{equation}
\label{beta-alpha-limits}
\beta_d(\alpha)\simeq
\begin{cases}
A_d\, \alpha^{-1} & \qquad \alpha\to 0,\\
B_d\, (\pi-\alpha)^{d-3} & \qquad \alpha \to \pi.
\end{cases}
\end{equation}
Inside a very thin cone, the first-passage process is very fast and
$\beta$ diverges. The proportionality constant is
$A_d=\tfrac{1}{2}\zeta(\delta)$ where $\zeta(\delta)$ is the first
positive zero of the Bessel\cite{jdw} function $J_\delta$ of order
$\delta = \tfrac{d-3}{2}$. For example, $A_3\cong 1.202412$,
$A_4=\frac{\pi}{2}\cong 1.570796$, and $A_5\cong 1.915852$.  As 
$\alpha \to \pi$, the exponent $\beta$ vanishes. The coefficient is
$B_d=\Gamma\big(\tfrac{d-2}{2}\big)/\big[2\,\sqrt{\pi}\,\Gamma\big(\tfrac{d-3}{2}\big)\big]$. The
corresponding behavior \eqref{beta-alpha-limits} holds only when
$d>3$. For $d=3$, the exponent vanishes logarithmically,
$\beta_3\simeq 1/[4\ln \frac{2}{\pi-\alpha}]$ (Fig.~\ref{fig-beta}).

\section{The cone approximation} 

\begin{figure}[t]
\centerline{\includegraphics[width=0.5\textwidth]{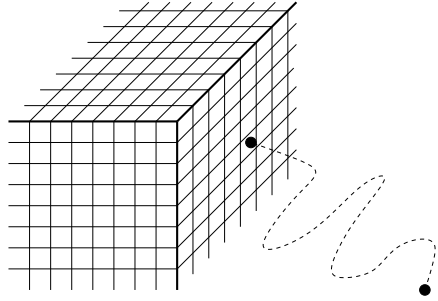}}
\caption{First-passage process outside a three-dimensional corner.}
\label{fig-corner}
\end{figure}

The above results for diffusion inside a circular cone can be used to
approximate diffusion in more complex geometries for which the
eigenvalues of the Laplace operator are not known analytically. We now
apply this approach to the exterior of the three-dimensional corner
(figure \ref{fig-corner}).  We are interested in the survival
probability $S(t)$ of a particle that is diffusing in the exterior of
an unbounded corner \cite{dr} in dimension three, and we that
the powerlaw decay \eqref{S-decay} holds, regardless of initial
position.

The corner geometry underlies a first-passage problem involving three
particles in one dimension. Let's assume that all particles start in
the positive half line, $0<x_3(0)<x_2(0)<x_1(0)$. Then, the
probability that the rightmost particle always remains in the positive
half line requires that the composite particle remains outside the
corner \cite{bk2}.  The complementary problem that the leftmost
particle remains in the positive half line is trivial: since all three
particles must remain in the positive half space, then $S(t)=[s(t)]^3$
with $s(t)\sim t^{-1/2}$ the survival probability in one
dimension. Hence, for diffusion inside the corner $\beta=3/2$.

We now replace the corner with an unbounded cone that has identical
solid angle. In $d$ dimensions, a cone with opening angle $\alpha$
occupies a fraction $V$ of the total solid angle,
\begin{equation}
\label{Valpha}
V(\alpha)=\frac{\int_0^\alpha d\theta\,(\sin\theta)^{d-2}} {\int_0^\pi
d\theta\,(\sin\theta)^{d-2}}.
\end{equation}
The normalization is such that $V(\pi)=1$, and of course,
$V(\pi/2)=1/2$. Equation \eqref{Valpha} follows from the Jacobian
$d\Omega/d\theta \propto(\sin\theta)^{N-2}$ with $\Omega$ being the
solid angle and $\theta$ the polar angle.

\begin{figure}[t]
\centerline{\includegraphics[width=0.5\textwidth]{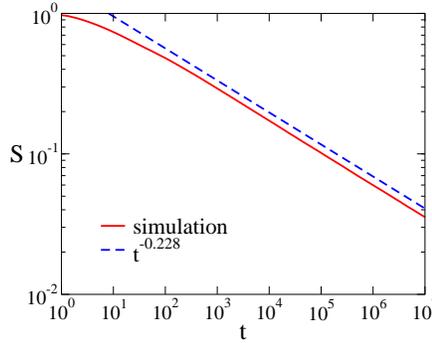}}
\caption{The survival probability in the exterior of a
  three-dimensional corner.  The solid line indicate results of
  numerical simulations and the dashed line, a best-fit.}
\label{fig-corner-mc}
\end{figure}

By substituting $d=3$ and $V=7/8$ into \eqref{Valpha}, we find that the
equivalent cone has opening angle $\alpha=2.418858$ and from equation
\eqref{beta3} we obtain 
\begin{equation}
\label{beta-corner}
\beta^{\rm cone}=0.216785\qquad {\rm whereas}\qquad \beta=0.228.
\end{equation}
The latter value was obtained from numerical simulations (figure
\ref{fig-corner-mc}).  Clearly, the cone approximation produces a
useful estimate for $\beta$.

The value $\beta^{\rm cone}$ is a lower bound. This assertion follows
from the fact that a sphere has the smallest eigenvalue of the Laplace
operator amongst the domains with the same volume. In two dimensions,
this result was conjectured by Rayleigh \cite{jwsr} and later proved
by Faber and Krahn\cite{ch}. In higher dimensions, the
Rayleigh-Faber-Krahn theorem remains valid \cite{ic}. This result
implies for example that the ground state of a quantum particle in a
box is lowest when the box is spherical; another consequence is that
the lifetime of a Brownian particle inside an absorbing box of a fixed
volume is highest if the box is spherical.  In our situation, we study
a composite particle in a general unbounded conical domain in $d$
dimensions. (By definition, any cone $\mathcal{R}$ with an apex at the
origin has the property that every ray from the origin to any point
inside $\mathcal{R}$ belongs to $\mathcal{R}$.)  The eigenvalue
problem generalizing Eqs.~\eqref{psi-eq-2} and \eqref{psi-theta-eq}
arising for circular cones is specified by
\begin{equation}
\label{LB}
\nabla^2 \psi = -2\beta(2\beta +d-2) \psi, \qquad 
\psi|_{\partial (S^{d-1}\cap \mathcal{R})} = 0.
\end{equation}
Here $\nabla^2$ is the {\em angular} component of the Laplacian. The
choice of the smallest eigenvalue ensures that the eigenfunction
$\psi$ is positive inside $\mathcal{R}$. We need the smallest
eigenvalue in the ``spherical cap'' $S^{d-1}\cap \mathcal{R}$, the
intersection of the cone $\mathcal{R}$ with unit sphere $S^{d-1}$, and
$\nabla^2$ is actually the Laplace-Beltrami operator on the spherical
cap. Fortunately, the Rayleigh-Faber-Krahn theorem generalizes to the
smallest eigenvalue of the Laplace-Beltrami operator on Riemannian
manifolds \cite{ic} and it proves that the smallest eigenvalue
$2\beta(2\beta +d-2)$, and therefore the smallest $\beta$, amongst all
spherical caps with the same volume (i.e., the same solid angle)
corresponds to the circular spherical cap.  Hence, the cone
approximation produces a lower bound.

Equation \eqref{beta-y-scaling} shows that for $d\gg 1$, there is
universal scaling behavior in terms of the variable
$y=(\cos\alpha)\sqrt{d}$. The same scaling variable also underlies the
fraction of space $V$ enclosed by the cone. The infinitesimal solid
volume element $d\Omega \propto (\sin\theta)^{d-2}d\theta $ in
\eqref{Valpha} becomes Gaussian, \hbox{$d\Omega \propto
\exp(-y^2/2)dy$}. Consequently, the volume which depends on the opening
angle and the dimension, $V\equiv V(\alpha,d)$, becomes a function of
a single variable $y$ 
\begin{equation}
\label{Vscaling}
V(\alpha,d)=\frac{1}{2}\,{\rm erfc}\left(\frac{y}{2}\right).
\end{equation}
Here ${\rm erfc}(\xi)=\big(2/\sqrt{\pi}\big)\int_\xi^\infty
e^{-u^2}du$ is the complementary error function. Combining
\eqref{Vscaling} and \eqref{scaling2}, we find $\beta$ as an implicit
function of volume $V$:
\begin{equation}
\label{beta-V}
D_{2\beta}\left(\sqrt{2}\,{\rm erfc}^{-1}(2V)\right)=0.
\end{equation}
Remarkably, this expression provides an exact description for the two
ordering problems we are interested in.

\section{Rank Statistics} 

\begin{figure}[t]
\centerline{\includegraphics[width=0.4\textwidth]{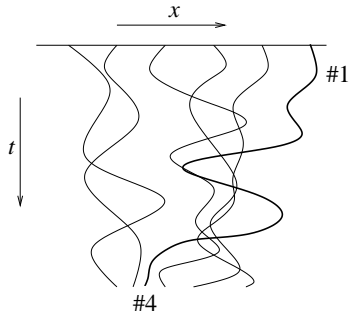}}
\caption{Illustration of the rank problem. Shown is the spacetime
  diagram of six Brownian particles in one dimension. The particles
  are ranked according position, from right to left, and in this
  example, the rank of the leader falls from one to four.}
\label{fig-rank}
\end{figure}

We now consider $N$ particles, each undergoing independent Brownian
motion in one dimension. The particles are ordered from right to left
with the rightmost particle at time $t=0$ regarded as the original
leader. As described in the introduction, we are interested in the
probability $R_n(t)$ that the rank of the original leader does not
fall below $n$ up to time $t$ (figure \ref{fig-rank}).  The extremal
cases $n=1$ and $n=N-1$ were considered previously
\cite{kr,bjmkr,bg,bmr,abk}, yet the long-time kinetics have been
established\cite{book} only for $N\leq 3$. 

\begin{figure}[t]
\centerline{\includegraphics[width=0.5\textwidth]{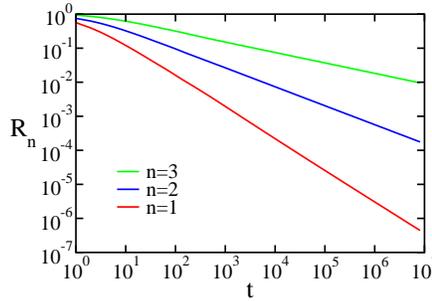}}
\caption{The survival probability $R_n$ versus rank $n$ for four
particles. The results are obtained from $10^{11}$, $10^9$, and $10^7$
independent Monte Carlo runs for $n=1$, $n=2$,and $n=3$ respectively.}
\label{fig-four}
\end{figure}

Based on the behavior for $N=3$ where the set
$\{\gamma_1,\gamma_2\}=\{3/4,3/8\}$ characterizes the decay $R_n\sim
t^{-\gamma_n}$, we expect $N-1$ distinct exponents,
\begin{equation}
\gamma_1>\gamma_2>\cdots>\gamma_{N-1}.
\end{equation}
Simulation results for $N=4$ confirm this behavior and give
(Fig.~\ref{fig-four}):
\begin{equation}
\label{gamma3}
\gamma_1=0.913, \qquad \gamma_2=0.556, \qquad \gamma_3=0.306.
\end{equation}
For $N=4$, the largest exponent has been calculated using
electrostatics $\gamma_1=0.91342$ \cite{bjmkr}.

As discussed above, the trajectories of $N$ particles in one dimension
can be represented by that of a compound particle in $N$
dimensions. The compound particle is confined to a volume $V_n$ of
space in which the position of the initial leader is larger than those
of at least $N-n$ of the particles.  Figure \ref{fig-pie} shows that
$V_1=1/3$ and $V_2=2/3$ for $N=3$. Generally
\begin{equation}
\label{Vn}
V_n=\frac{n}{N}.
\end{equation} 
In $N$ dimensions, the $N(N-1)/2$ hyperplanes $x_i=x_j$ with $i\neq j$
divide space into $N!$ equal ``chambers'' each with distinct particle
order (figure \ref{fig-wedge}). For example, the region
$x_1>x_2>\cdots>x_{N-1}>x_N$ is one such chamber where the initial
particle order is preserved, and this chamber occupies a fraction
$1/N!$ of the total solid angle. Since there are $(N-1)!$ chambers in
which the rank of the original leader equals $1$, then
$V_1=1/N$. There are $(N-1)!$ additional chambers in which the rank of
the leader equals $2$, and hence $V_2=2/N$. This argument establishes
\eqref{Vn}.

According to \eqref{Vn}, the compound particle is confined to a region
of space that is the union of $n(N-1)!$ chambers.  To apply the cone
approximation, we replace this complex region with an unbounded
spherical cone that encloses a fraction $V_n$ of space. We set $d=N-1$
so that this approximation is exact when $N=3$. By substituting
\eqref{Vn} and $d=N-1$ into \eqref{Valpha}, we obtain the opening
angle $\alpha$. The first-passage exponent is obtained as root of the
associated Legendre function specified in \eqref{betad} with
$d=N-1$. For example, when $N=3$, we obtain
\begin{equation}
\label{gamma3-cone}
\gamma^{\rm cone}_1=0.888644, \qquad \gamma_2=1/2, \qquad \gamma_3=0.300754.
\end{equation} 
These values, which represent formal lower bounds, provide useful
estimates for the numerically measured values \eqref{gamma3}.  Table 1
lists the largest and the smallest first-passage exponents for $3\leq
N\leq 10$. Although cone approximation slightly deteriorates as the
number of particle increases, overall the estimates remain faithful.

The predictions of the cone approximation in the limit $N\to\infty$
follow immediately from equations \eqref{Vn} and \eqref{beta-V}, 
\begin{equation}
\label{scaling-gamma}
D_{2\gamma}\left(\sqrt{2}\,{\rm erfc}^{-1}(2n/N)\right)=0.
\end{equation}
The exponent $\gamma$ which depends on two variables, $n$ and $N$,
becomes a universal function of the scaling variable $x=n/N$ as stated
in equation \eqref{scaling2}. Remarkably, the scaling function
specified by equation \eqref{scaling-gamma} is exact! Our numerical
simulations which use a remarkably large number of particles (up to
$N=10^4$) reveal that: (i) the first-passage exponent is a function of
the scaling variable $x$, and (ii) the scaling function
\eqref{scaling-gamma} is exact (figure \ref{fig-gamma}).

\begin{table}[t]
\begin{center}
  \tbl{The largest and the smallest exponents $\gamma_1$ and
    $\gamma_{N-1}$ for $3\leq N\leq 10$.  Simulation results are
    compared with the outcome of the cone approximation.}  {\small
\begin{tabular}{|c|l|l|l|l|}
\hline
N&$\gamma_1^{\rm cone}$&$\gamma_1$& $\gamma_{N-1}^{\rm cone}$&$\gamma_{N-1}$  \\
\hline
$3$&$3/4$&$3/4$&$3/8$&$3/8$\\
$4$&$0.888644$&$0.91$&$0.300754$&$0.306$\\
$5$&$0.986694$&$1.02$&$0.253371$&$0.265$\\
$6$&$1.062297$&$1.11$&$0.220490$&$0.234$\\
$7$&$1.123652$&$1.19$&$0.196216$&$0.212$\\
$8$&$1.175189$&$1.27$&$0.177469$&$0.190$\\
$9$&$1.219569$&$1.33$&$0.162496$&$0.178$\\
$10$&$1.258510$&$1.37$&$0.150221$&$0.165$\\
\hline
\end{tabular}}
\end{center}
\end{table}

This finding may indicate that in very high dimensions, the complex
boundary confining the composite particle approaches a limiting shape,
and that this limiting shape is conical. Another possible scenario is
that the complex boundary, which is formed by the intersection of
multiple hyperplanes, does not necessarily approach a limiting
conical shape, yet the deviations do not affect lowest eigenvalue of
the Laplace operator.

The scaling function \eqref{scaling-gamma} manifests that there is a
continuous spectrum  of first-passage exponents $0<\gamma<\infty$ when
$N\to\infty$. The smallest exponent vanishes as $N^{-1}$,
while the largest diverges logarithmically. More precisely,
\begin{equation}
\label{gamman-limits}
\gamma_n\simeq
\begin{cases}
\frac{1}{4}\ln N&\qquad n =1,\\
\frac{1}{N}\ln N & \qquad n={N-1}.
\end{cases}
\end{equation}
These limiting behaviors follow from asymptotic analysis of
\eqref{scaling-gamma} with \hbox{$n=1$} and $n=N-1$, and in both
cases, we recover results obtained using heuristic scaling arguments
\cite{kr,bjmkr}. We also note that median exponent, which
characterizes the probability that the leader always ranks above the
median particle, approaches the simple limit $\gamma_{N/2}\to 1/2$.

Knowledge of the scaling form \eqref{scaling-gamma}, merely the fact
that $\gamma$ is function of the variable $x=n/N$ is quite
powerful. This realization enables measurement of first-passage
exponents for up to $N=10^4$ particles, corresponding to diffusion in
a staggering spatial dimension $d=10^4$.  Whereas accurate measurement
of the (vanishing) smallest or the (diverging) largest exponent become
prohibitive \cite{bg} already at $N\approx 10$, we are able to compute
$\gamma$ by measuring the probability that the rank of the original
leader remains in the {\em first percentile} rather than strictly
remaining {\em first}.  The numerical simulation also show that not
only does the first-passage exponent becomes a function of the scaling
variable $x=n/N$, but so does the entire survival probability:
$R_n(N,t)\to \Phi(x,t)$ with $x=n/N$ in the limit $N\to\infty$.

\section{Inversion Statistics} 

\begin{figure}[t]
\centerline{\includegraphics[width=0.4\textwidth]{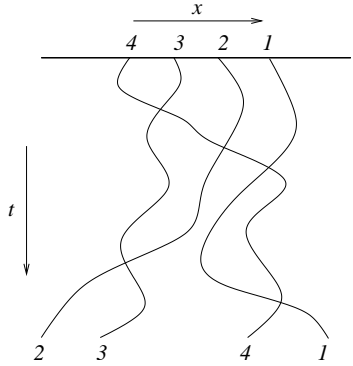}}
\caption{Illustration of the inversion problem.  Shown is the
spacetime diagram of four particles in one dimension. The number of
inversions grows from $k=0$ to $k=3$.}
\label{fig-inversions}
\end{figure}

Rank compares the location of a single particle, the initial leader,
with those of the rest of the particles. Generally, there are
$N(N-1)/2$ distinct pairs, and the number of pair inversion $k$ treats
all particles equally. If initially the particle locations are such
that \hbox{$x_N(0)<x_{N-1}(0)<\cdots<x_1(0)$}, then the number of pair
inversions is
\begin{equation}
\label{inversion}
k(t)=\sum_{i\neq j} \Theta\big(x_i(t)-x_j(t)\big),
\end{equation}
with $\Theta(x)$ the step function: $\Theta(x)=1$ when $x>0$ and
$\Theta(x)=0$ otherwise. In figure \ref{fig-inversions} the initial
order is $4321$ and later, when the order becomes $2341$, the number
of inversions equals $k=3$ as the pairs $23$, $24$, and $34$ are
inverted with respect to the initial configuration. The number of
inversions (also known as the Mahonian number) is a basic measure of a
permutation, and it is routinely used in sorting and ranking
algorithms \cite{wf,pam,gea,mb}. The number of inversions lies between
$0$ and $N(N-1)/2$.  

There are $N!$ different permutations of particle order and since the
diffusion process is ergodic, all of these permutations become equally
probable. It is easy to show that if initially the particles are
regularly spaced, the ``equilibration'' time is quadratic in $N$
\cite{eb}. Consequently, the equilibrium distribution of the inversion
number is given by the well-known Mahonian probability distribution
$P_k(N)$ that a random permutation of $N$ integers has $k$
inversions. The generating function reads \cite{dek}
\begin{equation}
\label{generating}
\sum_k P_k(N)s^k=\frac{1}{N!}\prod_{n=1}^N (1+s+s^2+\cdots + s^{n-1}).
\end{equation}
From this generating function, we can obtain the distribution
function.  For example, for two particles, $P_0(2)=P_1(2)=1/2$, and
for three particles, $P_0(3)=P_2(3)=1/6$ while $P_1(3)=P_2(3)=1/3$
(figure \ref{fig-six}b). We are particularly interested in the
cumulative distribution $V_k(N)$ that the number of inversions is
smaller than $k$
\begin{equation}
\label{cumulative}
V_k(N)=\sum_{j< k}P_j(N).
\end{equation}

From the generating function \eqref{generating}, we can obtain the
average number of inversions $\langle k\rangle$ and the standard
deviation $\sigma$:
\begin{equation}
\label{av-var}
\langle k\rangle= \frac{N(N-1)}{4} \qquad {\rm and}\qquad 
\sigma^2=\frac{N(N-1)(2N+5)}{72}\,.
\end{equation}
Notably, the standard deviation is large, $\sigma \sim
  N^{3/2}$.  When the number of particles is large, the distribution
  $P_k$ becomes normal and is characterized by the mean and the
  standard deviation \cite{wf,cjz}. Consequently, the cumulative
  probability $V_k(N)$ that the inversion number is $\leq k$
  approaches the complementary error function in the $N\to\infty$
  limit:
\begin{equation}
\label{normal}
V_k(N)\to {\rm erfc}\left(-\frac{w}{\sqrt{2}}\right)
\qquad {\rm with}\qquad w=\frac{k-\langle k\rangle}{\sigma}\,.
\end{equation} 

Let $M_k(t)$ be the survival probability that the number of inversions
remains smaller than $k$ up to time $t$.  We anticipate
\begin{equation}
\label{M-decay}
M_k\sim t^{-\nu_k}\qquad\text{with}\qquad \nu_1 >  \cdots >\nu_{N(N-1)/2}.
\end{equation}
Equation \eqref{nu3} shows that
$\{\nu_1,\nu_2,\nu_3\}=\{3/2,1/2,3/10\}$ when $N=3$. Numerical
simulations confirm that the number of distinct exponents equals
$N(N-1)/2$. In contrast with the rank statistics above, the number of
distinct exponents is quadratic in $N$.

\begin{table}[t]
\begin{center}
  \tbl{First-passage exponents for four particles. Results of Monte
    Carlo simulations, $\nu_k$ versus the outcome of the
    cone approximation, $\nu_k^{\rm cone}$.}
  {\small
\begin{tabular}{|c|c|c|c|c|c|c|}
\hline $k$&$1$&$2$&$3$&$4$&$5$&$6$\\ 
\hline
$\nu_k^{\rm cone}$& $2.67100$ & $1.17208$ & $0.64975$ & $0.39047$ & $0.24517$ 
& $0.14988$  \\
\hline 
$\nu_k$& $3$ & $1.39$ & $0.839$ & $0.455$ & $0.275$ & $0.160$ \\ 
\hline
\end{tabular}}
\end{center}
\end{table}

The quantity $M_1$ equals the probability that the particle order
remains intact, and the respective exponent $\nu_1$ is known
analytically \cite{mef,hf,fg,gz,djg,ck}
\begin{equation}
\label{nu1}
\nu_1=\frac{N(N-1)}{4}.
\end{equation}
When $k=1$, the trajectory of the compound Brownian particle is
confined to one chamber, $x_1>x_2>\cdots>x_N$.

We again use the cone approximation to estimate the first-passage
exponents. The fraction \eqref{cumulative} calculated from the
generating function \eqref{generating} is substituted into
\eqref{Valpha} to give the opening angle $\alpha$. Using this angle
and $d=N-1$, we deduce the exponent from equation
\eqref{betad}. Results of the cone approximation for the case $N=4$
are listed in table 2. Again, we obtain useful estimates for the
first-passage exponents.

\begin{figure}[t]
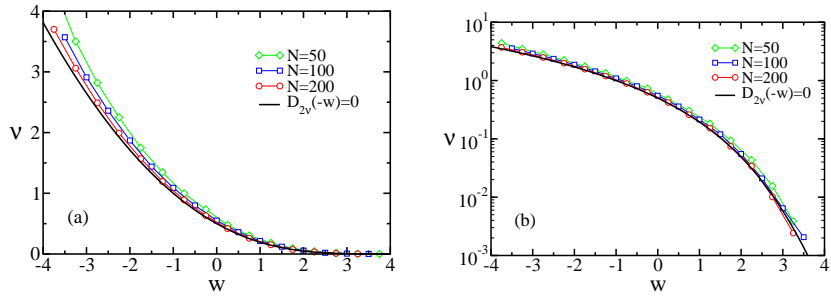

\centerline{\includegraphics[width=0.45\textwidth]{fig13a}\hspace{.2in}
\includegraphics[width=0.45\textwidth]{fig13b}}
\caption{The exponent $\nu$ versus the scaling variable
$w$.  Predictions of the cone approximation are compared with results of
Monte Carlo simulations with $N=50$, $N=100$, and $N=200$ particles.}
\label{fig-nu}
\end{figure}

To obtain the limiting behavior when the number of particles diverges,
$N\to\infty$, we substitute \eqref{normal} into the
scaling relation \eqref{scaling2}. The scaling exponent $\nu$ becomes
a function of the scaling variable $w$ defined in \eqref{normal} 
and is root of the parabolic cylinder equation, 
\begin{equation}
\label{scalingw}
D_{2\nu}(-w)=0.
\end{equation}
Hence, if we focus on the probability that the number of inversions
does not exceed a fixed number $w$ of standard deviations from the
mean, the first-passage process becomes independent of the number of
particles. This realization is especially useful in the case of
inversion statistics: the family of exponents is quadratic in $N$, and
furthermore, the scaling variable $w$ guides us to measure $\nu_k$ in
a scaling window of size $N^{3/2}$ around the average, $\langle
k\rangle\simeq N^2/4$.  Indeed, by using a finite number of
measurements, we can verify that $\nu\equiv \nu_k(N)$ becomes a
function of the scaling variable $w$.  Again, the cone approximation
predicts the exact scaling function governing $\nu$ (figure
\ref{fig-nu}).

\section{Conclusions}

We studied first-passage properties of a Brownian particle in
unbounded circular cones. The first-passage exponent characterizing
the decay of the survival probability is root of a transcendental
equation involving the associated Legendre equation. Explicit
expressions for this exponent are feasible only in two and four
dimensions.  Another simplification occurs when $d\gg 1$ where the
exponent, which generally depends on the cone opening angle $\alpha$
and $d$, becomes a function of a single variable
$\sqrt{d}\,\cos\alpha$. This universal function is obtained
analytically as a root of the parabolic cylinder function.

We analyzed two first-passage problems involving the order of an
ensemble of independent Brownian particles in one dimension. The
trajectories of the $N$ Brownian particles define a compound
trajectory of a single Brownian particle in $N$ dimensions, more
precisely in an infinite cone with numerous plane boundaries. As an
approximation, we replaced this cone with complex boundary by a
circular cone having an identical solid angle.  We showed that the
cone approximation yields a lower bound and surprisingly good
approximations for the first-passage exponents.  

There are families of first-passage exponents that govern kinetics of
ordering of Brownian particles or equivalently, kinetics of
first-passage in high dimensions. We discussed only two out of many
possible measures for particle order: the rank of the original leader
and the number of pair inversions. In one case, the number of
first-passage exponents is linear in the number of particles; in the
other case, it is quadratic. The discrete spectrum of exponents
becomes continuous when the number of particles
diverges. Surprisingly, the cone approximation predicts this spectrum
exactly.

An important challenge is to establish when does the confining
geometry becomes equivalent to a cone in a high-dimensional space and
the conditions for the cone approximation to be asymptotically
exact. We presented two cases for which the scaling function is
predicted exactly, but there are counterexamples where the cone
approximation correctly predicts the scaling form but fails to predict
the actual scaling function \cite{bk2}.

The exponents exhibit scaling when the number of particles
diverges. This realization is relevant for numerical simulations
\cite{pg,obdkgs} where it is difficult to measure exponents that are
either much smaller than or much larger than unity. The scaling
variable specifies the region where the first-passage exponents are of
order one. For example, for the rank problem the scaling behavior
implies that the rank should be measured in percentiles.

First-passage processes are inherently nonequilibrium. By definition
the particle trajectory is restricted to certain regions in space
whereas other regions are strictly excluded. For example, in the
corner geometry \eqref{fig-corner} an unrestricted Brownian particle
visits all eight ``quadrants'' with equal probabilities $p=1/8$. Yet,
for first-passage processes inside a three-dimensional corner, we have
$p=1$ inside the corner and $p=0$ otherwise. By exploiting the cone
geometry, we have shown that the equilibrium probabilities \eqref{Vn}
and \eqref{normal} translate into knowledge of nonequilibrium
properties, and in particular, first-passage exponents. Interestingly,
geometrical properties allow us to deduce exact nonequilibrium
properties of a particle system from its equilibrium properties. It
will be interesting to extend these results to other geometries
\cite{ls,kr1} and to establish connections with other problems
involving ordering of Brownian particles.

\bigskip\noindent 
We acknowledge DOE grant DE-AC52-06NA25396 for support.

\end{document}